\documentstyle{elsart}
\newbox\mybox
\newcommand\fverb{\setbox\mybox=\hbox\bgroup\verb}
\newcommand\fverbdo{\egroup\medskip\noindent\fbox{\unhbox\mybox}\ }
\newcommand\fverbit{\egroup\item[\fbox{\unhbox\mybox}]}

\newcommand\init[1]{\setbox\mybox=\hbox{{\beeg #1}~}%
                   \noindent\global\hangindent=\wd\mybox\global\hangafter-2%
                   \sc\smash{\llap {\lower 13.2pt \box\mybox}}}
\def\v1{\vspace{1cm}}
\def\be{\begin{equation}}
\def\ee{\end{equation}}
\def\bc{\begin{center}}
\def\ec{\end{center}}
\def\vh{\varphi}

\newcommand{\bea}{\begin{eqnarray}}
\newcommand{\eea}{\end{eqnarray}}
\renewcommand{\thesection}{\arabic{section}.}
\renewcommand{\thesubsection}{\thesection\arabic{subsection}.}
\begin{document}
\begin{frontmatter}
\title{ Hamiltonian Cosmological Perturbation Theory}
\author[dub]{B.M. Barbashov,}
\author[dub]{V.N. Pervushin,}
\author[china,dub,itep,asc]{A.F. Zakharov,}
\author[dub]{V.A. Zinchuk}
\address[dub]{Bogoliubov Laboratory for Theoretical Physics, JINR, 141980 Dubna,
Russia}
\address[china]{National Astronomical Observatories of Chinese Academy of
Sciences, 20A Datun Road, Chaoyang District, Beijing, 100012, China}
\address[itep]{Institute of Theoretical and Experimental Physics, B. Cheremushkinskaya, 25,
117259, Moscow, Russia}
\address[asc]{Astro Space Center of Lebedev Physics Institute of RAS, Moscow}
\begin{abstract}
The Hamiltonian approach to cosmological perturbations in general
relativity  in finite space-time
 is developed, where a cosmological scale factor  is identified with
 spatial averaging the metric determinant logarithm.
 This identification preserves the number of variables and
   leads to a  cosmological perturbation theory with
 the scalar potential perturbations in contrast to the kinetic
  perturbations in  the Lifshitz version which
are responsible for the ``primordial power spectrum'' of CMB in
the inflationary model. The Hamiltonian approach enables to
explain this ``spectrum''  in terms of scale-invariant variables
and to consider other topical problem of modern cosmology  in the
context of   quantum cosmological creation of both universes and
particles from the stable Bogoliubov vacuum.

\end{abstract}
\begin{keyword}
General Relativity and Gravitation, Cosmology, Observational Cosmology
\\[2mm]
{\sc PACS}: 95.30.Sf, 98.80.-k, 98.80.Es
\end{keyword}
\end{frontmatter}
%

\section{Introduction}

The cosmological perturbation theory in general relativity (GR)
\cite{lif,bard} based on the separation of the cosmological scale
factor by the transformation
$g_{\mu\nu}=a^2\widetilde{g}_{\mu\nu}$
 is  one of the basic tools applied for analysis of modern
observational data including including Cosmic Microwave Background
(CMB).

In the present paper
 we discuss the problem of the relation between
 the cosmological perturbation theory  and
 the Hamiltonian approach \cite{dir,ADM} to GR,
 where a similar scale factor  was considered in \cite{pp}
 as the  homogeneous invariant evolution parameter in accordance with the
Hamiltonian diffeomorphism subgroup $x^0 \to
\widetilde{x}^0=\widetilde{x}^0(x^0)$ \cite{vlad}  meaning in fact
that the coordinate evolution parameter  $x^0$ is not observable.
The statement of the problem is to formulate the cosmological
perturbation theory in terms of diffeo-invariant quantities.

The content of the paper
 is the following.
 In Section 2, the statement of the problem is given.
 In Section 3, it is shown that the separation of the scale
 factor can lead to exact resolution of the energy constraint in GR and to
   its Hamiltonian reduction. Sections
  4 and 5 are devoted to  cosmological models of both  classical and quantum
  universes
  that follows from the reduced theory.
    The Hamiltonian perturbation
  theory and its comparison with Lifshitz's one are given in Section  6.

\section{\label{s-2} Statement of problem}

 GR is given in terms of metric components
 and fields  $f$ by the Hilbert  action
 \be\label{gr}
 S=\int d^4x\sqrt{-g}\left[-\frac{\vh_0^2}{6}R(g)
 +{\cal L}_{(\rm M)}(\vh_0|g,f)\right]
 \ee
 and the  space-time geometric interval
 $ds^2=g_{\mu\nu}dx^\mu
 dx^\nu$, where the parameter $
 \varphi_0=\sqrt{3/{8\pi G_0}}=\sqrt{3M^2_{\rm
 Planck}/{8\pi}}
 $
 scales all masses, and $G_0$ is
 the Newton coupling constant in units $\hbar=c=1$.
The Hamiltonian approach is formulated by means of
   a geometric interval
 \be\label{adm-f}
 ds^2=g_{\mu\nu}dx^\mu dx^\nu\equiv\omega_{(0)}\omega_{(0)}-
 \omega_{(1)}\omega_{(1)}-\omega_{(2)}\omega_{(2)}-\omega_{(3)}\omega_{(3)},
 \ee where
   $\omega_{(\alpha)}$ are  linear differential forms
  \cite{fock29}  in
 terms of the Dirac variables \cite{dir}
\be \label{adm}
 \omega_{(0)}=\psi^6N_{d}dx^0,~~~~~~~~~~~
 \omega_{(b)}=\psi^2 {\bf e}_{(b)i}(dx^i+N^i dx^0);
 \ee
 here $\psi$ is the spatial metrics determinant,
  ${\bf e}_{(a)i}$ are triads with $\det |{\bf
 e}|=1$, $N_{ d}$ is the Dirac lapse function, and $N^i$ is the shift vector.
 The comparison of this  interval with the one
 \be\label{li} ds^2=a^2(\eta)\left[(1+2\Phi)d\eta^2-2N_kdx^kd\eta
 -(1-2\Psi)(dx^k)^2-dx^idx^j(h_{ij})\right] \ee
 used in
 the cosmological perturbation theory \cite{lif}
  raises to the following question: Is it possible
  to formulate the Hamiltonian approach
  to GR in terms of the metric components (\ref{adm}) so that
  the conformal time $\eta$ in Eq. (\ref{li})
  as the measurable one of a cosmic photons
  becomes diffeo-invariant quantity
 with respect to the time-coordinate transformations
  $x^0 \to \widetilde{x}^0=\widetilde{x}^0(x^0)$, and
        the potential $\Psi$ does not
        contain one more a homogeneous component, in order to preserve the number
        of variables of GR?

\section{\label{s-3}Separation of  Scale Factor and  Hamiltonian Reduction}

 The invariance of the action (\ref{gr})
 and interval  (\ref{adm-f}) expressed in terms of the Fock --
 Dirac simplex (\ref{adm})
 with respect to
  time-coordinate transformations  $x^0 \to \widetilde{x}^0=\widetilde{x}^0(x^0)$
 means  that  a diffeo-invariant
 ``evolution parameter'' in GR coincides with one of homogeneous
 variables \cite{pp,vlad,Y}. The cosmological evolution is
 the irrefutable observational
 argument in favor of
 existence of such a homogeneous variable  considered in GR
  as the cosmological scale factor.
  The  cosmological
 scale factor  $a(x_0)$ introduced
 by the scale transformation:
 $g_{\mu\nu}=a^2(x_0)\widetilde{g}_{\mu\nu}$, where $\widetilde{g}_{\mu\nu}$
 is defined by (\ref{adm-f}) and (\ref{adm}), where
 $\widetilde{N}_d=a^{2}{N}_d$ and
 $\widetilde{\psi}^2=a^{-1}\!\psi^2.$
 In  order to keep the number of variables of GR, the scale factor
can be defined using the spatial averaging $ \log
\sqrt{a}\equiv\langle{\log {\psi}}\rangle
 \equiv V_0^{-1}\int d^3x \log {\psi}$, so that the rest
 scalar component $\widetilde{\psi}$
  satisfies  the identity
\be\label{non1}
 \langle\log\widetilde{\psi}\rangle
 \equiv V_0^{-1}\int d^3x \log\widetilde{\psi}
 =
 V_0^{-1}\int d^3x \left[\log{\psi}
 -\left\langle{ \log{\psi}}\right\rangle\right]\equiv 0,
 \ee
 where $V_0=\int d^3x$ is finite volume.
 The similar scale transformation of a curvature
 $
 \sqrt{-g}R(g)=a^2\sqrt{-\widetilde{g}}R(\widetilde{g})-6a
 \partial_0\left[{\partial_0a}\sqrt{-\widetilde{g}}~\widetilde{g}^{00}\right]$
  converts action (\ref{gr}) into
 \be\label{1gr}
 S=\widetilde{S}-
 \int\limits_{V_0} dx^0 (\partial_0\vh)^2\int {d^3x}{\widetilde{N}_d}^{-1},
 \ee
 where $\widetilde{S}$
  is the action (\ref{gr})  in
 terms of metrics $\widetilde{g}$ and
 the running scale  of all masses
 $\vh(x^0)=\vh_0a(x^0)$   and $(\widetilde{N}_d)^{-1}=
 \sqrt{-\widetilde{g}}~\widetilde{g}^{00}$.
 One can construct the Hamiltonian function using
 the definition of a set of  canonical
 momenta:
\bea \label{pph}
 P_\vh&=&\frac{\partial L}{\partial (\partial_0\vh)}
 = -2V_0\partial_0\vh
 \left\langle(\widetilde{N}_d)^{-1}\right\rangle=
 -2V_0\frac{d\varphi}{d\zeta}\equiv-
2V_0 \vh',
 \\
 \label{gauge}
 p_{\psi}&=&\frac{\partial {\cal L}}{\partial (\partial_0\log\widetilde{\psi})}\equiv
 -\frac{4\vh^2}{3}\cdot\frac{\partial_l(\widetilde{\psi}^{6}N^l)-
 \partial_0(\widetilde{\psi}^{6})}{\widetilde{\psi}^{6}\widetilde{N_d}},
 \eea
where $d\zeta=\langle(\widetilde{N}_d)^{-1}\rangle^{-1}dx^0$ is a
time-interval invariant with respect to
  time-coordinate transformations
  $x^0 \to \widetilde{x}^0=\widetilde{x}^0(x^0)$.
One can construct the  Hamiltonian form of the action  in terms of
momenta $P_\vh$ and $P_{ F}=[{p_{\psi}}, p^i_{{(b)}},p_f]$
including (\ref{pph}), (\ref{gauge})
 \be\label{hf}
S=\int dx^0\left[\int d^3x \left(\sum_F P_F\!\partial_0
F\!+\!C\!-\!\widetilde{N}_d\widetilde{T}^0_0\right)\!-\!P_{\varphi}\partial_0\varphi+
\frac{P_{\varphi}^2}{4\int dx^3 ({\widetilde{N}_d})^{-1}}\right],
\ee where
 ${\cal C}=N^i {T}^0_{i} +C_0p_{\psi}+ C_{(b)}\partial_k{\bf
e}^k_{(b)}$
  is the sum of constraints
  with the Lagrangian multipliers $N^i,C_0,~C_{(b)}$ and the energy--momentum tensor
  components $T^0_i$; these constraints include
   the transversality  $\partial_i {\bf e}^{i}_{(b)}\simeq 0$ and the Dirac
 minimal  surface \cite{dir}:
 \be\label{hg}
{p_{\psi}}\simeq 0 ~~~~\Rightarrow ~~~~
\partial_j(\widetilde{\psi}^6{\cal N}^j)=(\widetilde{\psi}^6)'
~~~~~ ({\cal N}^j=N^j\langle \widetilde{N}_d^{-1}\rangle).
 \ee
The explicit dependence of $\widetilde{T}_0^0$ on
$\overline{\psi}$
  can be given in terms of the scale-invariant  Lichnerowicz
  variables  \cite{Y}
  $\omega^{(L)}_{(\mu)}=\psi^{-2}\omega_{(\mu)}$:
 \be\label{t00}
 \widetilde{T}^0_0= \widetilde{\psi}^{7}\hat \triangle \widetilde{\psi}+
  \sum_I \widetilde{\psi}^Ia^{I/2-2}\tau_I, \ee
   where $\hat \triangle
 F\equiv({4\varphi^2}/{3})\partial_{(b)}\partial_{(b)}F$ is
 the Laplace operator and
  $\tau_I$ is partial energy density
  marked by the index $I$ running a set of values
   $I=0$ (stiff), 4 (radiation), 6 (mass), 8 (curvature), 12
   ($\Lambda$-term)
in accordance with a type of matter field contributions, and $a$
is the scale factor.

 The
 energy constraint ${\delta S[\vh_0]}/{\delta  \widetilde{N_d}}=0$ takes the
 algebraic form
 \be\label{nph}
 -
 \frac{\delta \widetilde{S}[\vh]}{\delta  \widetilde{N}_d}
 \equiv \widetilde{T}^0_0=\frac{(\partial_0\varphi)^2}{\widetilde{N}_d^2}
 =\frac{P_\vh^2}{4V_0^2[{\langle(\widetilde{N}_d)^{-1} \rangle
 \widetilde{N}_d}]^2},
 \ee
 where $T^0_0$ is the local energy density by definition.
The spatial averaging
  of this equation multiplied by $\widetilde{N}_d$ looks like the energy constraint
 \be \label{ec}
 ~~~~~~~~~~~~~~~~~~~~~~~~~~~~~~ P^2_{\varphi}=E^2_{\varphi},
 \ee
 where the  Hamiltonian functional $ E_\vh=2\int
 d^3x(\widetilde{T}^0_0)^{1/2}= 2V_0{\langle
 (\widetilde{T}^0_0)^{1/2}\rangle}
 $
 can be treated as the ``universe energy'' by analogy with the ``particle energy'' in
 special relativity (SR).
 Eqs. (\ref{pph}) and (\ref{ec}) have the exact solution
 \be\label{13c}
 \zeta(\varphi_0|\varphi)
 \equiv\int dx^0 {\left\langle
 (\widetilde{N}_d)^{-1}\right\rangle}^{-1}
 =\pm
 \int_{\vh}^{\vh_0}
 {d\widetilde{\vh}}{{\langle
 (\widetilde{T}^0_0(\widetilde{\vh}))^{1/2}\rangle}}^{-1}
 \ee
 well known as the Hubble-type evolution.
 The local part of Eq. (\ref{nph}) determines
 a   part of
 the Dirac lapse function invariant with respect to
 diffeomorphisms  of the
Hamiltonian formulation $x^0 \to
\widetilde{x}^0=\widetilde{x}^0(x^0)$ \cite{vlad}:
 \be\label{13ec}
 N_{\rm inv}={\langle(\widetilde{N}_d)^{-1} \rangle
 \widetilde{N}_d}={{\left\langle\sqrt{{\widetilde{T}^0_0}}\right\rangle}}
 \left(\sqrt{{\widetilde{T}^0_0}}\right)^{-1}.
 \ee

 One can find
 evolution of all field variables $F(\vh,x^i)$  with respect to
 $\vh$ by the variation of the ``reduced'' action obtained as
   values of the Hamiltonian form of the initial action  (\ref{hf}) onto
 the energy constraint  (\ref{ec}) \cite{pp}:
 \be\label{2ha2} S|_{P_\vh=\pm E_\vh} =
 \int\limits_{\vh}^{\vh_0}d\widetilde{\vh} \left\{\int d^3x
 \left[\sum\limits_{  F}P_{  F}\partial_\vh F
 +\bar{\cal C}\mp2\sqrt{\widetilde{T}_0^0(\widetilde{\vh})}\right]\right\},
\ee
 where $\bar{\cal C}={\cal
 C}/\partial_0\widetilde{\vh}$.  The reduced Hamiltonian
 $\sqrt{\widetilde{T}_0^0}$ is
 Hermitian,
 if the  minimal surface
 constraint
 (\ref{hg}) removes a negative
 contribution of $p_{\psi}$ from the energy density \cite{pl}.
 The reduced action (\ref{2ha2}) shows us
 that the initial data at the beginning $\vh=\vh_I$ are independent of
  the present-day ones at  $\vh=\vh_0$,
  therefore
  the proposal about an existence of the  Planck epoch $\vh=\vh_0$
   at the beginning \cite{bard} looks
  very doubtful. Let us consider consequences of
  the classical reduced theory (\ref{2ha2}) and quantization of
  the energy constraint (\ref{ec}) without the ``Planck epoch''
  at the beginning.

\section{\label{s-4}Observational Data in  Terms of Scale-Invariant Variables}

 Let us assume that the local density
 $T_0^0=\rho_{(0)}(\vh)+T_{\rm f}$
 contains a tremendous  cosmological background
$\rho_{(0)}(\vh)$.
 The  low-energy decomposition
  of ``reduced''  action (\ref{2ha2})  $2 d\vh \sqrt{\widetilde{T}_0^0}= 2d\vh
\sqrt{\rho_{(0)}+T_{\rm f}}
 =
 d\vh
 \left[2\sqrt{\rho_{(0)}}+
 T_{\rm f}/{\sqrt{\rho_{(0)}}}\right]+...$
 over
 field density $T_{\rm f}$ gives the sum
 $S|_{P_\vh=+E_\vh}=S^{(+)}_{\rm cosmic}+S^{(+)}_{\rm
 field}+\ldots$, where the first  term of this sum
 $S^{(+)}_{\rm cosmic}= +
 2V_0\int\limits_{\vh_I}^{\vh_0}\!
 d\vh\!\sqrt{\rho_{(0)}(\vh)}$ is  the reduced  cosmological
 action,
 whereas the second one is
  the standard field action of GR and SM
 \be\label{12h5} S^{(+)}_{\rm field}=
 \int\limits_{\zeta_I}^{\zeta_0} d\zeta\int d^3x
 \left[\sum\limits_{ F}P_{ F}\partial_\eta F
 +\bar{{\cal C}}-T_{\rm f} \right]
 \ee
 in the  space determined by the interval
 \be\label{d2}
 ds^2=d\zeta^2-[e_{(a)i}(dx^i+{\cal N}^id\zeta]^2;
 ~~\partial_ie^i_{(a)}=0,~~\partial_i{\cal N}^i=0
 \ee
 with  conformal time
 $d\eta=d\zeta=d\vh/\rho_{(0)}^{1/2}$ as the diffeo-invariant
 and scale-invariant quantity, coordinate distance
 $r=|x|$,
 and running masses
 $m(\zeta)=a(\zeta)m_0$.
 We see that
  the  correspondence principle leads to the theory (\ref{12h5}),
  where the scale-invariant conformal  variables and coordinates are
    identified  with observable ones
    and the cosmic evolution with the evolution of masses:
  $$
\frac{E_{\rm emission}}{E_0}=\frac{m_{\rm atom}(\eta_0-r)}{m_{\rm
atom}(\eta_0)}=\frac{\vh(\eta_0-r)}{\vh_0}=a(\eta_0-r)
=\frac{1}{1+z}.
$$
The conformal observable distance  $r$ loses the factor $a$, in
comparison with the nonconformal one $R=ar$. Therefore, in this
 case, the redshift --
  coordinate-distance relation $d\eta=d\vh/\sqrt{\rho_0(\vh)}$
  corresponds to a different
  equation
  of state than in the standard one  \cite{039,Danilo}.
    The best fit to the data  including
  cosmological SN observations \cite{sn} 
 requires a cosmological constant $\Omega_{\Lambda}=0.7$,
$\Omega_{\rm CDM}=0.3$ in the case of the Friedmann
``scale-variant quantities`` of standard cosmology, whereas for
the ``scale-invariant conformal
 quantities''
 these data are consistent with  the dominance of the stiff state
of free scalar field $\Omega_{\rm Stiff}=0.85\pm 0.15$,
$\Omega_{\rm CDM}=0.15\pm 0.10$ \cite{039}. If $\Omega_{\rm
Stiff}=1$, we have the square root dependence of the scale factor
on conformal time $a(\eta)=\sqrt{1+2H_0(\eta-\eta_0)}$. Just this
time dependence of the scale factor on
 the measurable time (here -- conformal one) is used for description of
 the primordial nucleosynthesis \cite{Danilo,three}.
Thus the stiff state can describe all epochs including the
creation of a quantum universe.

\section{\label{s-5}The Quantum Universe}
  \label{ch-2}

 We have seen above that in the ``reduced''  action (\ref{2ha2}) momenta
 ${P_\vh}_{\pm}=\pm E_\vh$
   become  the generators of evolution of all variables with respect to the
  evolution parameter $\vh$ \cite{pp} forward and backward,
 respectively. The negative energy problem can be solved by
 the primary quantization of the energy constraint
 $[{P^2_\vh}-E^2_\vh]\Psi_{\rm u}=0$ and the secondary quantization
 $\Psi_{\rm u}=(1/\sqrt{2E_\vh)}[A^++A^-]$ by the Bogoliubov
 transformation $ A^+=\alpha
 B^+\!+\!\beta^*B^-$, in order to diagonalize the equations of
 motion by the condensation of ``universes''
 $<0|\frac{i}{2}[A^+A^+-A^-A^-]|0>=R(\vh)$
 and describe  cosmological creation of a  ``number'' of universes
  $<0|A^+A^-|0>=N(\vh)$
  from the stable Bogoliubov vacuum  $B^-|0>=0$.
 The vacuum postulate $B^-|0>=0$ leads to an arrow of the conformal
time $\eta\geq 0$ and its absolute point of reference $\eta= 0$ at
the moment of creation $\vh=\vh_I$ \cite{pp}.
 The cosmological creation of the
 {\it ``universes''} is described by the Bogoliubov equations \cite{origin}
\bea\nonumber
 \frac{dN}{d\vh}=\frac{dE_\varphi}{2E_\varphi d\vh}
 \sqrt{4N(N+1)-R^2} \label{nu},~~~
 \frac{dR}{d\vh}=-{2E_\vh}
 \sqrt{4N(N+1)-R^2} \label{ru}.
 \eea
In   the model of the stiff  state $\rho=p$, where
 $E_\vh=Q/\vh$, these  equations
 have an  solution  \cite{origin}
$
 N(\vh)=\frac1{4Q}R(\vh)=\frac{1}{4Q^2-1}
 \sin^2\left[\sqrt{Q^2-\frac{1}{4}}~~\ln\frac{\vh}{\vh_I}\right]\not
 =0,
$
 where
 $
 \vh=\vh_I\sqrt{1+2H_I\eta}
 $. 
  The initial data
 $\vh_I=\vh(\eta=0),H_I=\vh'_I/\vh_I=Q/(2V_0\vh_I^2)$ are considered  as a
 point of creation or annihilation of a universe; whereas the Planck value of
 the running mass scale $\vh_0=\vh(\eta=\eta_0)$ belongs to the present
 day moment $\eta_0$.

 These initial data $\vh_I$ and $H_I$ are determined by
 parameters of matter cosmologically created from the Bogoliubov
 vacuum  at the beginning of a universe $\eta\simeq 0$.
 In the Standard
 Model (SM),
   W-, Z- vector bosons have maximal probability of this
 cosmological creation
 due to their mass singularity~\cite{114:a}.
 The uncertainty principle
  $\triangle E\cdot\triangle \eta \geq 1$
  (where $\triangle E=2M_I,\triangle \eta=1/2H_I$)
  shows that at
   the moment of creation of  vector bosons their  Compton
   length
 defined by its inverse mass
 $M^{-1}_{\rm I}=(a_{\rm I} M_{\rm W})^{-1}$ is close to the
 universe horizon defined in the
 stiff state as
 $H_{\rm I}^{-1}=a^2_{\rm I} (H_{0})^{-1}$.
 Equating these quantities $M_{\rm I}=H_{\rm I}$
 one can estimate the initial data of the scale factor
 $a_{\rm I}^2=(H_0/M_{\rm W})^{2/3}=10^{-29}$ and the Hubble parameter
 $H_{\rm I}=10^{29}H_0\sim 1~{\rm mm}^{-1}\sim 3 K$.
 Just at this moment there is  an effect of intensive
  cosmological creation of the vector bosons described in  \cite{114:a};
 in particular, the distribution functions of the longitudinal   vector bosons
demonstrate a large contribution of relativistic momenta. Their
temperature $T_c$  can be estimated from the equation in the
kinetic theory for the time of establishment of this temperature $
\eta^{-1}_{relaxation}\sim n(T_c)\times \sigma \sim H $, where
$n(T_c)\sim T_c^3$ and $\sigma \sim 1/M_I^2$ is the cross-section.
This kinetic equation and values of the initial data $M_{\rm I} =
H_{\rm I}$ give the temperature of relativistic bosons $
 T_c\sim (M_{\rm I}^2H_{\rm I})^{1/3}=(M_0^2H_0)^{1/3}\sim 3 K
$ as a conserved number of cosmic evolution compatible with the SN
data \cite{039}.
 We can see that
this  value is surprisingly close to the observed temperature of
the CMB radiation
 $ T_c=T_{\rm CMB}= 2.73~{\rm K}$. The primordial mesons before
 their decays polarize the Dirac fermion vacuum and give the
 baryon asymmetry frozen by the CP -- violation
 so that $n_b/n_\gamma \sim X_{CP} \sim 10^{-9}$,
 $\Omega_b \sim \alpha_{\rm \tiny QED}/\sin^2\theta_{\rm Weinberg}\sim
 0.03$, and $\Omega_R\sim 10^{-5}\div 10^{-4}$~\cite{114:a}.
 All these results
 testify to that all  visible matter can be a product of
 decays of primordial bosons, and the observational data on CMB
 can reflect  parameters of the primordial bosons, but not the
 matter at the time of recombination.
 The length of  the semi-circle on the surface of  the last emission of
photons at the life-time
  of W-bosons
  in terms of the length of an emitter
 (i.e.
 $M^{-1}_W(\eta_L)=(\alpha_W/2)^{1/3}(T_c)^{-1}$) is
 $\pi \cdot 2/\alpha_W$.
 It is close to $l_{min}\sim  210 $ of CMB,
 whereas $(\bigtriangleup T/T)$ is proportional to the inverse number of
emitters~
 $(\alpha_W)^3 \sim    10^{-5}$.
 The temperature history of the expanding universe
 in this case looks like the
 history of evolution of masses of elementary particles in the cold
 universe with the constant conformal temperature $T_c=a(\eta)T=2.73~ {\rm K}$
 of the cosmic microwave background.
The equations describing the longitudinal vector bosons
 in SM, in this case, are close to
 the equations that  are  used, in  the
 inflationary model \cite{bard}, for
 description of the ``power primordial spectrum'' of the CMB radiation.

\section{\label{s-6}The Potential Perturbations and Shift Vector}

In order to simplify equations of the scalar potentials
  ${N}_{\rm inv},\widetilde{\psi}$, one can introduce a new
 table of symbols:
 $
 N_{\rm s}=\psi^7 N_{\rm inv}, ~\widetilde{T}=
 \sum_{I} \widetilde{\psi}^{(I-7)}
 a^{\frac{I}{2}-2}\tau_I,~~\rho_{(0)}=\langle
(\widetilde{T}^0_0)^{1/2}\rangle^2=\vh'^2
 $.
 The variations of the action (\ref{hf}) with respect to
 $N_{\rm s},\log\widetilde{\psi}$ lead to equations
 \bea\label{4-1}
\hat \triangle
 \widetilde{\psi}+\widetilde{T}&=&\frac{\widetilde{\psi}~{}^7
 \rho_{(0)}}{N^2_{\rm s}},
 \\\label{4-2}
\widetilde{\psi}\hat \triangle{N}_{\rm s}
 +{N}_{\rm s}
 \frac{d\widetilde{T}}{d\log\widetilde{\psi}}
 +7\frac{\widetilde{\psi}~{}^7
 \rho_{(0)}}{N_{\rm s}}&=&\rho_{(1)},
 \eea
 respectively, where $\rho_{(1)}=\langle\widetilde{\psi}\hat \triangle{N}_{\rm s}
 +{N}_{\rm s}\widetilde{\psi}\partial_{\widetilde{\psi}}{\widetilde{T}}
 +7{\widetilde{\psi}~{}^7
 \rho_{(0)}}/{N_{\rm s}}\rangle$.

 For $N_{\rm s}=1-{\nu}_1$ and
 $\widetilde{\psi}=1+{\mu}_1$ and the small deviations $\mu_1,\nu_1\ll 1$ the
 first orders of Eqs.  (\ref{4-1}) and
(\ref{4-2}) take the form
 \bea\label{e1-2}
   [-\hat{\triangle}+14\rho_{(0)}-\rho_{(1)}]\mu_{1}~~ +&
   2\rho_{(0)}\nu_1&=~~~\overline{\tau}_{(0)},
 \\\label{ec1-2}
 [7\cdot 14\rho_{(0)}\!-\!14\rho_{(1)}+\!\rho_{(2)}]\mu_1~~
 +&[-\hat{\triangle}+
14\rho_{(0)}\!-\!\rho_{(1)}]\nu_1&=7\overline{\tau}_{(0)}-\overline{\tau}_{(1)},
 \eea
where
 $
 \rho_{(n)}=\langle\tau_{(n)}
 \rangle\equiv\sum_II^na^{\frac{I}{2}-2}\langle\tau_{I}\rangle
 $.
This choice of variables
 determines $\widetilde{\psi}=1+{{\mu_1}}$
 and $N_{\rm s}=1-{{\nu_1}}$ in the form of a sum
  \bea\label{12-17}
 \widetilde{\psi}&=&1+\frac{1}{2}\int d^3y\left[D_{(+)}(x,y)
\overline{T}_{(+)}^{(\mu)}(y)+
 D_{(-)}(x,y) \overline{T}^{(\mu)}_{(-)}(y)\right],\\\label{12-18}
 N_{\rm s}&=&1-\frac{1}{2}\int d^3y\left[D_{(+)}(x,y)
\overline{T}^{(\nu)}_{(+)}(y)+
 D_{(-)}(x,y) \overline{T}^{(\nu)}_{(-)}(y)\right],
  \eea
 where $\beta$ is given by
 \be\label{beta}
 \beta=\sqrt{1+[\langle \tau_{(2)}\rangle-14\langle
 \tau_{(1)}\rangle]/(98\langle \tau_{(0)}\rangle)},
 \ee
 \be\label{1cur1}\overline{T}^{(\mu)}_{(\pm)}=\overline{\tau_{(0)}}\mp7\beta
  [7\overline{\tau_{(0)}}-\overline{\tau_{(1)}}],
 ~~~~~~~
 \overline{T}^{(\nu)}_{(\pm)}=[7\overline{\tau_{(0)}}-\overline{\tau_{(1)}}]
 \pm(14\beta)^{-1}\overline{\tau_{(0)}}
 \ee
 are the local currents, $D_{(\pm)}(x,y)$ are the Green functions satisfying
 the equations
 \bea\label{2-19}
 [\pm \hat m^2_{(\pm)}-\hat \triangle
 ]D_{(\pm)}(x,y)=\delta^3(x-y),
 \eea
 where $\hat m^2_{(\pm)}= 14 (\beta\pm 1)\langle \tau_{(0)}\rangle \mp
\langle \tau_{(1)}\rangle$.

  In the case of point mass distribution in a finite volume $V_0$ with the
zero pressure
  and  the  density
  $\overline{\tau_{(0)}}(x)=\overline{\tau_{(1)}}(x)
/6\equiv  M\left[\delta^3(x-y)-{1}/{V_0}\right]$,
 solutions   (\ref{12-17}),  (\ref{12-18}) take
 a  form
 \bea\label{12-21}
  \widetilde{\psi}&=1+\mu_1=1+
  \frac{r_{g}}{4r}\left[{\gamma_1}e^{-m_{(+)}(z)
 r}+ (1-\gamma_1)\cos{m_{(-)}(z)
 r}\right],\\\label{12-22}
 N_{\rm s}&=1-\nu_1=1-
 \frac{r_{g}}{4r}\left[(1-\gamma_2)e^{-m_{(+)}(z)
 r}+ {\gamma_2}\cos{m_{(-)}(z)
 r}\right],
 \eea
 where
 $
  {\gamma_1}=\frac{1+7\beta}{2},~~~
 {\gamma_2}=\frac{14\beta-1}{28\beta},~~
 r_{g}=\frac{3M}{4\pi\vh^2},~~
 r=|x-y|,~~m^2_{(\pm)}=\frac{3\hat m^2_{(\pm)}}{4\vh^2}.
 $

 The minimal surface (\ref{hf})
  $\partial_i[\overline{\psi}^6{\cal N}^i]-(\overline{\psi}^6)'=0$
 gives the shift of the coordinate
  origin in the process of evolution
 \be \label{2-23}
{\cal
 N}^i=\left(\frac{x^i}{r}\right)\left(\frac{\partial_\zeta V}{\partial_r V}\right),~~~
 ~~~V(\zeta,r)=\int\limits_{}^{r}d\widetilde{r}
 ~\widetilde{r}^2\widetilde{\psi}^6(\zeta,\widetilde{r}).
  \ee
Solutions (\ref{12-21}),  (\ref{12-22}) have spatial oscillations
and the nonzero shift of the coordinate
  origin of the type of
  (\ref{2-23}).
In the infinite volume limit $\langle \tau_{(n)}\rangle=0,~a=1$
 solutions (\ref{12-21}) and  (\ref{12-22}) coincide with
 the isotropic version of  the Schwarzschild solutions:
 $\widetilde{\psi}=1+\frac{r_g}{4r}$,~
 ${N_{\rm s}}=1-\frac{r_g}{4r}$,~$N^k=0$.

 Now one can
  compare the Hamiltonian perturbation theory  with  the
 standard cosmological perturbation theory (\ref{li})
 \cite{lif}
 where $\overline{\Phi}=\nu_1+\mu_1$ ,
 $\overline{\Psi}=2\mu_1$, $N^i=0$.
   We note that the zero-Fourier harmonics
 of the spatial determinant is taken into account in \cite{bard}
 twice that
 is an obstruction to the Hamiltonian method. The Hamiltonian
 approach
 shows us that if this double counting is removed, the equations
 of  scalar potentials  (\ref{e1-2}), and
  (\ref{ec1-2}))
 will not contain time derivatives that are responsible for
the CMB ``primordial power spectrum'' in the inflationary model
\cite{bard}. However,  these    equations of the Lifshits
perturbation theory  are close to ones of the primordial vector
bosons. We have seen above that the Hamiltonian approach to GR gives
us another possibility to explain the CMB ``spectrum'' by
cosmological creation of vector W-, Z- bosons.

 The next differences are a nonzero shift vector and the spatial oscillations of
  the scalar potentials determined by $ m^2_{(-)}$.
 In the diffeo-invariant version of cosmology \cite{039}, the
  SN data dominance of stiff state  determines the parameter
  of spatial oscillations
  $ m^2_{(-)}=\frac{6}{7}H_0^2[\Omega_{\rm R}(z+1)^2+\frac{9}{2}\Omega_{\rm
  Mass}(z+1)]$. The redshifts in the recombination
  epoch $z_r\sim 1100$ and the clustering parameter
 $
 r_{\rm clustering}={\pi}/{ m_{(-)} }\sim {\pi}/[{
 H_0\Omega_R^{1/2} (1+z_r)}] \sim 130\, {\rm Mpc}
 $
  recently
 discovered in the researches of large scale periodicity in redshift
 distribution \cite{da}
 lead to a reasonable value of the radiation-type density
  $10^{-4}<\Omega_R\sim 3\cdot 10^{-3}$ at the time of this
  epoch.

\vspace{1cm}

{\bf Acknowledgements}\\
 The authors are grateful to D. Blaschke,  A.~Gusev, A.~Efremov,
  E.~Kuraev,
 V.~Priezzhev,  and  S.~Vinitsky
 for fruitful discussions. AFZ is grateful to
 the National Natural Science Foundation of China
(NNSFC) (Grant \# 10233050)  for a partial financial support.

\end{document}